\documentclass[12pt]{article}
\usepackage{epsfig}
\begin{document}
%set bwid 6; set fwid 6; set hwid 6; set pwid 6; set xwid 8; set ywid 6;
%igset lwid 6

\begin {center}
{\Large \bf An interesting feature of BES III data for
$J/\Psi \to \gamma (\eta' \pi \pi )$}
\vskip 5mm
{D V Bugg \footnote{email: d.bugg@rl.ac.uk}\\
{\normalsize  \it Queen Mary, University of London, London E1\,4NS,
UK} \\
[3mm]}
\end {center}
\date{\today}

\begin{abstract}
%\noindent
The $\eta (1835)$ is confirmed clearly in new BESIII data for
$J/\Psi \to \gamma (\eta '\pi \pi$); the angular distribution
of the photon is consistent with a pseudoscalar.
This makes it a candidate for an $s\bar s$ radial excitation of $\eta
'$ and $\eta (1440)$ (or one or both of $\eta (1405)$ and $\eta
(1475)$).
However, a conspicuous feature of the BES III data is the absence of
evidence for $\eta (1440) \to \eta ' \pi \pi$ while it is well known
that $\eta (1440)$ appears in $\eta \pi \pi$.
Can these facts be reconciled? There is in fact a simple explanation.
The channel $\eta (1440) \to \eta \pi \pi$ may be explained by the two-step
process $\eta (1440) \to [K^*K]_{L=1}$ and $[\kappa \bar K ]_{L=0}$,
followed by $K\bar K \to a_0(980) \to \eta \pi$.
This process does not produce any significant $\eta '\pi$ signal
because of the Adler zero close to the $\eta '\pi$ threshold.

\end{abstract}
New BESIII data confirm $\eta (1835)$ in
$J/\Psi \to \gamma (\eta ' \pi \pi )$ \cite {BESIII}.
Huang and Zhu point out that $\eta (1835)$ may be explained naturally
as an $s\bar s$ radial excitation of $\eta '(958)$ and $\eta (1440)$
\cite {Huang}.
However, a conspicuous feature of the BESIII data is the
absence of $\eta (1440) \to \eta '\pi \pi$, although there is a
small and narrow peak attributed to $f_1(1510) \to \eta ' \pi \pi$.
The data contrast strongly with the clear $\eta (1440)$ signal observed
in $J/\Psi \to \gamma (\eta \pi \pi )$ by Mark III \cite {Mark3}, DM2
\cite {DM2} and BES I \cite {BES1}.

%Fig. 1
\begin{figure}[htb]
\begin{center}
\vskip -43mm
\epsfig{file=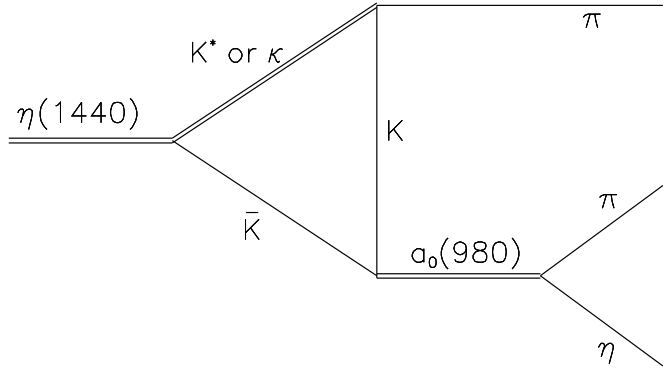,width=12cm}
\vskip -29mm
\caption {Triangle graph for $\eta (1440) \to \eta \pi \pi$}
\end{center}
\end{figure}
There is a straightforward explanation shown in Fig. 1.
The dominant signals attributed to $\eta (1440)$ are
decays to $K^*(890)\bar K$, $\kappa \bar K$ (where $\kappa$ is the
$K\pi$ S-wave), and a weak $\eta \pi \pi$ channel.
The $K\bar K$ pairs from the first two channels can rescatter through
$a_0(980)$ to $\eta \pi$.
Triangle graphs for this process were calculated by Anisovich et al. 
treating intermediate $K\bar K$ pairs as real particles 
\cite {Anisovich}.
This model fitted Crystal Barrel data on
$\bar pp \to \eta \pi^+\pi^-\pi^+\pi^-$ successfully.
A feature of those data is that the $\eta \pi$ peak is 30-50 MeV above
the $K\bar K$ threshold, as expected from the phase space for real
intermediate $K\bar K$ pairs.
For the $\eta '\pi \pi$ final state, $a_0(980) \to \eta '\pi$ is
attentuated strongly by the Adler zero at
$s = m^2_{\eta '} - m^2_\pi /2$, where $m$ are masses \cite {a0980}.

It will clearly be important to show that the small $\eta '\pi \pi$ peak
in BESIII data fitted to $f_1(1510)$ has $J^P = 1^+$.
It looks too narrow and too high in mass to be due to $\eta (1440)$.
It is also obviously important to look for the $\eta (1835)$ in 
$J/\Psi \to \gamma (K\bar K\pi)$, as a check on the $s\bar s$
assignment of Huang and Zhu.

An important general comment is that fits to $\eta (1405)$ and 
$\eta (1475)$ must include the full $s$-dependence of opening channels 
and associated dispersive corrections to resonance amplitudes.
Breit-Wigner amplitudes of constant width are seriously misleading.
As an example, the full formula for the production amplitude of
$\eta (1475) \to [K^*\bar K]_{L=1}$ is
\begin {eqnarray}
f &\propto & kF(k)B(k)/[M^2 - s - m(s) - ig^2\rho (k)F^2(k)B^2(k)] \\
m(s) &=& \frac {(s - M^2)}{\pi} {\rm P} \int 
\frac {M\Gamma _{total}(s') ds'}{(s' - s)(M^2-s')}.
\end {eqnarray}
The $\bar K K^*$ channel opens at a mass of $\sim 1385$ MeV and its phase
space $\rho$ rises as $k^3$, where $k$ is the momentum of $K$ and $K^*$
in the $\bar K K^*$ rest frame;
$B$ is the Blatt-Weisskopf centrifugal barrier factor 
$1/(1 + k^2R^2)^{1/2}$ and $g$ is a coupling constant.
The term $m(s)$ is the so-called `running mass' and is required to 
make the amplitude fully analytic; P stands for the principal value
integral. 
The integral includes a subtraction on resonance, making it strongly 
convergent.
The term $F(k)$ is a form factor. 
Fits to a large number of data sets with a Gaussian 
$F(k) = \exp (-\alpha k^2)$ give $\alpha = (2.25 \pm 0.25)$ (GeV/c)$^{-2}$, 
corresponing to a radius $R = 0.73$ fm for the meson cloud.
The width of the $K^*$ needs to be folded into the calculation, as does
the phase space for production processes such as 
$J/\Psi \to \gamma (\bar K K^*)$ or $\bar pp \to (\bar K K^*)\sigma$.  

%Fig. 2
\begin{figure}[htb]
\begin{center}
\vskip -12mm
\epsfig{file=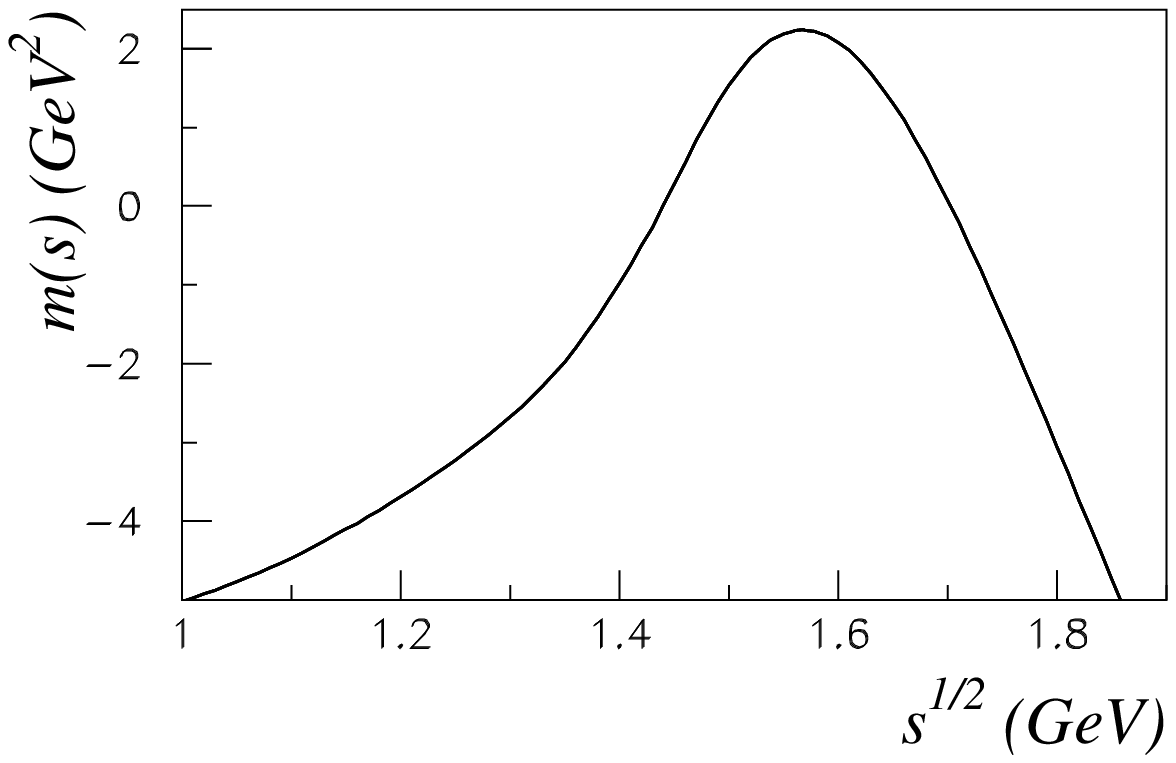,width=8cm}
\vskip -6mm
\caption {$m(s)$ for $\eta (1475) \to \bar K K^*$.}
\end{center}
\end{figure}
It turns out that the term $m(s)$, shown in Fig. (2), varies more 
rapidly than $(M^2 - s)$ near resonance, because of the $k^3$
dependence of $\rho (\bar K K^*)$.
As a result, a fit to the resonance of constant width and mass 1476
MeV quoted by the Particle Data Group \cite {PDG} moves the mass 
down to 1439 MeV. 
This raises doubts about the existence of two narrowly 
spaced $J^P = 0^-$ states.
The effect on $\eta (1405)$ is smaller, because the $\kappa K$
phase space varies as $k$.


\begin{thebibliography}{99}
\bibitem {BESIII}          %1
M. Ablikim et al., [BESIII Collaboration], arXiv: 1012.3510.
\bibitem {Huang}           %2
J. Huang and S-L. Zhu, Phys. Rev. D 73 (2006) 014023.
\bibitem {Mark3}           %3
J.Z. Bai et al., Phys. Rev. Lett. 65 (1990) 2507.
\bibitem {DM2}             %4
J.-E. Augustin et al., Phys. Rev. D 46 (1992) 1951.
\bibitem {BES1}            %5
J.Z. Bai et al., Phys. Lett. B 446 (1998) 356.
\bibitem {Anisovich}       %6
A.V. Anisovich {\it et al.}, Nucl. Phys. A 690 (2001) 567.
\bibitem {a0980}           %7
D.V. Bugg, Phys. Rev. D 78 (2008) 074023.
\bibitem {PDG}             %8
K. Nakamura et al., [Particle Data Group], J. Phys. G: Nucl.
Part. Phys. 37 (2010) 075021.
\end {thebibliography}
\end {document}